# Lattice modes in a dusty plasma crystal

A. Abdikian[1,a)] and Zahida Ehsan[2,b)]
[1]Department of Physics, Malayer University, Malayer, Iran
[2] INFN, Department of Physical Sciences, University of Naples 80126, Napoli, Italy and Università degli Studi del Molise, 86090 Pesche (IS), Italy
[3]COMSATS Institute of Information Technology (CIIT), Defence Road, Off Raiwind Road, Lahore, Pakistan.

**Abstract:** A model is presented to explain the normal mode features of dust particles in a planar zigzag crystal chain for the first and second neighbors. The degrees of freedom of particles are the longitudinal and transverse displacements in plane coupled by the first and second neighbor harmonic forces in two-dimensions (2D). The constant electric force duded to the electrodes to keep the zigzag structure is calculated. The coupling between transverse and longitudinal dust-lattice (DL) modes is derived. The latter is considered due to the energy of the electrostatic (Yukawa) potential.
Moreover coupled (acoustic and optical) and decoupled (longitudinal and transverse) branches of dust lattice modes for different lattice parameters and structures are studied.
 Propagation of the longitudinal and acoustic modes is found to be strictly dependent on the value of the distance between the two chains; below that value mode may not propagate
Finally it is shown that the frequencies of the acoustic (optical) branches increase (decrease) with increasing the distance between the two chains.

a) Electronic mail: abdykian@gmail.com
b) Electronic mail: Ehsan.zahida@gmail.com

## I. INTRODUCTION

Despite a history spanning nearly a century, research in complex  (dusty) plasmas advanced tremendously over the last two decades both academically and due to its variety of applications in space and modern astrophysics, semiconductor technology, fusion devices, plasma chemistry, etc. [1-9].

Some of the novel experimental discoveries that originated are dusty plasma crystals, dust Mach cones, dust voids [8-9] etc., whereas an idea of "dust atoms and molecules" was also put forward recently [10] and it was studied that in the localized region of the EM wave, the density of grains increases and the ions, following the grains, start clustering around them, while electrons are pushed away from that region, this scenario suggests the possibility of crystallization of dust atoms [11].

Coulomb clusters (or Wigner crystals) an important phenomenon in classical physics [12] is still recognized substantially by the physics community. They are formed due to the strong electrostatic interaction of charged particles captured within an external confining electrostatic potential. Examples of such systems include quantum dots [13], electrons or ions in Paul or Penning traps [14], ions in storage rings [15], electron droplets on liquid helium [16], or polymer particles in colloidal suspensions [17] etc.

Investigation in this area had been further encouraged in the past few decades with the serendipitous discovery of complex (dusty) plasma which provide an ideal and versatile system to study Coulomb (Yukawa) crystals; these are characterized when the average potential energy of the neighboring micron sized charged dust particles considered by Yukawa potential exceeds the kinetic or thermal energy [18,19]. This dusty plasma crystal also supports wave modes such as longitudinal and transverse dust lattice wave (DLW) [20].
Usually for the investigation of dusty crystals, plastic microspheres with diameter of 1 – 10 micrometer are trapped in a plasma discharge.
Also, by applying appropriate confinement potentials in laboratory, typical lattice configurations including simple one-dimensional (1D) arrangements, horizontal 2D layers, and 3D patterns (*bcc*, *fcc* and *hcp*) so-called Coulomb balls [18,19] have also been analyzed in great detail. However, transition between the various configurations had not been investigated until recently when Melzer presented experiments where a zigzag configuration has been generated in which the transition from the one-dimensional linear chain of particles to 2D structures has been done by increasing the particle number or by variation of the anisotropy of the confinement [21].

Furthermore, the dispersion relation of the longitudinal and transverse modes in these linear chains was also derived by Melzer from the normal mode spectra and was found to be in good agreement with the natural phonon spectrum in chains [21].

This configuration may help us to study the dynamical processes in a variety of condensed matter [22-25] and biophysical systems [26] in 2D and for 3D to study the helix structures (DNA, protein, etc.) [27-30].
For making a zigzag structure, the dust cluster is trapped in 3D cigar-shaped potential well, provided by the barrier on the electrode and is given as

$$\Phi(x,y,z) = \frac{1}{2} m \left( \omega_x^2 x^2 + \omega_y^2 y^2 + \omega_z^2 z^2 \right) \qquad (1)$$

where *x, y* are Cartesian coordinates in the horizontal plane and *z* denotes the vertical direction. The barrier distorts the electrostatic equipotential lines in the plasma sheath above the electrode and thus forms an anisotropic confinement since the size of the barrier in the *x*-direction is much larger than that in the *y*-direction [22]. The vertical confinement is due to gravity and electric field force and is much stronger than the horizontal confinement [31] with high vertical resonance frequency $\omega_z \gg \omega_x, \omega_y$. Consequently, the motion of the microspheres is restricted to *x* and *y* directions only. Hence, the vertical direction will be ignored in the following discussions. The total energy of N dust particles confined in the anisotropic horizontal confinement is then obtained by [30].

$$E = \frac{1}{2}m\omega_0^2 \sum_{i=1}^{N}(\beta x_i^2 + y_i^2) + \frac{z^2 e^2}{4\pi\varepsilon_0}\sum_{i>j}^{N}\frac{\exp(-r_{ij}/\lambda_D)}{r_{ij}} \qquad (2)$$

The first term is the potential energy in the confining well of the barrier, the second is the (screened) Coulomb energy between the particles. Here, we have chosen $\omega_0 = \omega_y$ and $\beta = \omega_x^2/\omega_y^2$ is the parameter of anisotropy. Due to the shape of the barrier, $\omega_x < \omega_y$ and thus $\beta < 1$. Isotropic 2D confinement corresponds to $\beta = 1$. With $\beta > 1$ the confinement is anisotropic towards a 1D shape [21].

In this work, we plan to investigate the behavior of long-chain molecules, such as hydrocarbons with zigzag structures and will derive analytically the dispersion relations in a 2D dust plasma crystal. Here the amount of electric force is predicted that maintain the zigzag structure in such a way that it can overcome the Yukawa electrostatic force between the particles. In doing so, the G (kappa) function has been introduced which helps us to determine if the melting occurs or not? Moreover, this method predicts whether the oscillation modes can propagate or not (it depends on the value of the distance between the two chains) and provides an appropriate physical description.

## II. MATHEMATICAL MODELLING

We consider a planar two-dimensional zigzag dust crystal and assume that the dust grains having a constant negative charge are located at equidistant sites "$a$", for the convenience we assume it to be an infinite crystal.

Since our aim is to derive motion equations for the crystalline monolayer, for that we consider two-dimensional motion in the horizontal (along the $x$ axis, longitudinal) and vertical (along the $y$ axis, transverse) directions from the equilibrium positions. Fig. (1) illustrates a schematic representation of the zigzag crystal consisting of two coupled linear chains directed along the $x$-axis with lattice spacing "$a$" and "$h$" as the distance between the two chains.

Four nearest particles with labels $n+1$, $n-1$, $n+2$ and $n-2$ can also be seen in the Fig. (1); where the distance between closest neighbors (located in $y=h$) and the central particle is $b = \sqrt{a^2/4 + h^2}$. If $h \leq \sqrt{3}/2a$ then it is viewed as a first neighboring particle otherwise (when $h > \sqrt{3}/2a$), it is considered to be a second neighbor. In other words, the central particle can form an equilateral or and isosceles triangle with its neighbors. When $h$ is small the particle getting close and a larger confining force must be used for overcoming the repulsive electrostatic force, i.e. the largest "$\beta$" must be chosen.

Assuming that the particle is the origin of the plane at site $n$, then the position of dust particles at equilibrium are $(a,0)$, $(-a,0)$, $(a/2,h)$, $(-a/2,h)$. If the particles are not at their equilibrium positions, we may define the four-length variables $l_1$, $l_2$, $l_3$, and $l_4$, which present the distances from the particle $n$ to the nearest particles, respectively.

If the particles are not at their equilibrium positions, we may define the four length variables representing the distances from central particle to the nearest neighboring particles, given below:

$$l_1 = \sqrt{(a/2 + u_{n+1} - u_n)^2 + (h - v_{n+1} - v_n)^2},  \quad (3)$$

$$l_2 = \sqrt{(a/2 + u_n - u_{n-1})^2 + (h - v_{n-1} - v_n)^2},  \quad (4)$$

$$l_3 = \sqrt{(a + u_{n+2} - u_n)^2 + (v_{n+2} - v_n)^2},  \quad (5)$$

$$l_4 = \sqrt{(a + u_n - u_{n-2})^2 + (v_n - v_{n+2})^2}.  \quad (6)$$

where $u_i$ and $v_i$ (for $i=1, 2, 3, 4$) stand for the displacements of the respective particles from their equilibrium positions in both $x$ and $y$ directions.

The electrostatic binary interaction force $F(r)$ exerted onto two dust grains situated at distance $r$ is derived from the potential function $U(r) = q^2 \exp(-r/\lambda_D)/(4\pi\varepsilon_0 r)$, where $r$ is the distance between two dust particles, $\lambda_D$ is the Debye radius and $\varepsilon_0$ denotes the electric susceptibility of vacuum, viz. $F(r) = -\partial U(r)/\partial r$. The interaction potential energy presents a minimum value, so the particles vibrate around the equilibrium. We may expand the potential energy around equilibrium at $r = b$ and $r = a$ for the first and second neighbors, viz.

$$U(r) = U(b) + (r-b)\left.\frac{\partial U}{\partial r}\right|_{r=b} + \frac{1}{2}(r-b)^2 \left.\frac{\partial^2 U}{\partial r^2}\right|_{r=b} + \cdots, \quad (7)$$

and

$$U(r) = U(a) + (r-a)\left.\frac{\partial U}{\partial r}\right|_{r=a} + \frac{1}{2}(r-a)^2 \left.\frac{\partial^2 U}{\partial r^2}\right|_{r=a} + \cdots,$$

By defining the "spring" constant $G_1 = \partial U/\partial r|_{r=b}$ and $G_2 = \partial^2 U/\partial r^2|_{r=b}$ for first and $G_1' = \partial U/\partial r|_{r=a}$ and $G_2' = \partial^2 U/\partial r^2|_{r=a}$ for second neighbors, respectively and setting the potential energy at equilibrium to zero, we obtain

$$U(r) \cong G_1(r-b) + G_1'(r-a) + \frac{1}{2}G_2(r-b)^2 + \frac{1}{2}G_2'(r-a)^2, \quad (8)$$

We calculate the polynomial coefficients $G_1$, $G_1'$, $G_2$ and $G_2'$, for a Yukawa system; then the corresponding expressions read as

$$G_1 = -\frac{Q^2}{4\pi\varepsilon_0 \lambda_D^2}\frac{1+\kappa}{\kappa^2}e^{-\kappa}, \qquad G_2 = \frac{Q^2}{4\pi\varepsilon_0 \lambda_D^3}\frac{2+2\kappa+\kappa^2}{\kappa^3}e^{-\kappa},$$

$$G_1' = -\frac{Q^2}{4\pi\varepsilon_0 \lambda_D^2}\frac{1+\kappa'}{\kappa'^2}e^{-\kappa'}, \qquad G_2' = \frac{Q^2}{4\pi\varepsilon_0 \lambda_D^3}\frac{2+2\kappa'+\kappa'^2}{\kappa'^3}e^{-\kappa'}. \quad (9)$$

where $\kappa = b/\lambda_D$ and $\kappa' = a/\lambda_D$ are the dimensionless and positive real numbers of lattice parameter; also $\kappa = (b/a)\kappa'$ in other words $\kappa$ is a function of $h$).

The components of the linear force exerted on the particle $n$ by the nearest particles, in the $x$ and $y$ directions, are given by

$$F_x^{(int)} = -G_1\left(-\frac{a/2+u_{n+1}-u_n}{l_1}+\frac{a/2-u_{n-1}+u_n}{l_2}\right) - G_1'\left(-\frac{a+u_{n+2}-u_n}{l_3}+\frac{a-u_{n-2}+u_n}{l_4}\right) \quad (10)$$

$$-G_2\left(-\frac{a/2+u_{n+1}-u_n}{l_1}(l_1-b)+\frac{a/2-u_{n-1}+u_n}{l_2}(l_2-b)\right)$$

$$-G_2'\left(-\frac{a+u_{n+2}-u_n}{l_3}(l_3-a)+\frac{a-u_{n-2}+u_n}{l_4}(l_4-a)\right)$$

and

$$F_y^{(int)} = -G_1\left(-\frac{h-v_{n+1}-u_n}{l_1}-\frac{h-v_{n-1}-v_n}{l_2}\right) - G_1'\left(\frac{v_n-v_{n+2}}{l_3}+\frac{v_n-v_{n-2}}{l_4}\right) \quad (11)$$

$$-G_2\left(-\frac{h-v_{n+1}-u_n}{l_1}(l_1-b)-\frac{h-v_{n-1}-v_n}{l_2}(l_2-b)\right) - G_2'\left(\frac{v_n-v_{n+2}}{l_3}(l_3-a)+\frac{v_n-v_{n-2}}{l_4}(l_4-a)\right)$$

For the small amplitude waves, i.e., $\Delta u_i \ll a$ and $\Delta v_i \ll a$, we may neglect the nonlinear terms, then we are left with following force equations:

$$F_x^{(int)} = -\left(\frac{h^2 G_1}{b^3}+\frac{(b^2-h^2)G_2}{b^2}\right)(2u_n-u_{n+1}-u_{n-1})$$
$$+\frac{ah}{2b^2}\left(\frac{G_1}{b}-G_2\right)(v_{n+1}-v_{n-1}) - G_2'(2u_n-u_{n+2}-u_{n-2}) \quad (12)$$

and

$$F_y^{(int)} = \frac{2hG_1}{b} - \left(\frac{a^2 G_1}{4b^3}+\frac{(4b^2-a^2)G_2}{4b^2}\right)(2v_n+v_{n+1}+v_{n-1})$$
$$+\frac{ah}{2b^2}(\frac{G_1}{b}-G_2)(u_{n-1}-u_{n+1})-\frac{G_1'}{a}(2v_n-v_{n+2}-v_{n-2}) \quad (13)$$

Thus performing the standard, straightforward algebraic steps, we obtain the linearized set of motion equations for the nth dust particle in the zigzag chain in the $x$ and $y$ directions.

$$M\frac{\partial^2 u_n}{\partial t^2} = F_x^{(int)}+F_x^e = -\left(\frac{h^2 G_1}{b^3}+\frac{(b^2-h^2)G_2}{b^2}\right)(2u_n-u_{n+1}-u_{n-1})$$
$$+\frac{ah}{2b^2}\left(\frac{G_1}{b}-G_2\right)(v_{n+1}-v_{n-1}) - G_2'(2u_n-u_{n+2}-u_{n-2}) - M\omega_x^2 u_n \quad (14)$$

and

$$M\frac{\partial^2 v_n}{\partial t^2} = F_y^{(int)} + F_y^e = \frac{2hG_1}{b} - \left(\frac{a^2 G_1}{4b^3} + \frac{(4b^2 - a^2)G_2}{4b^2}\right)(2v_n + v_{n+1} + v_{n-1})$$
$$+ \frac{ah}{2b^2}(\frac{G_1}{b} - G_2)(u_{n-1} - u_{n+1}) - \frac{G_1'}{a}(2v_n - v_{n+2} - v_{n-2}) - QE - M\omega_y^2 v_n \quad (15)$$

where $M$ represents the mass of the dust particles. On the other hand, assuming a smooth and continuous electric field in the $y$ direction and the dust charge $Q$ near the equilibrium position $z_0$, we can expand the electric confinement at equilibrium, as follows

$$F_{ele}(y) = QE \approx \frac{2h}{b}G_1$$

Here to compensate the gravity force, a special electric force is used to suspend the dust particles in the vertical direction. In this situation for maintaining a zigzag structure there must be an amount of electric force equals to $(2h/b)G_1$ in the y direction so that it can overcome the Yukawa electrostatic force between the particles.

So, the net force which acts on $n$th dust particle from the others in the $y$ direction is

$$M\frac{\partial^2 v_n}{\partial t^2} = -\left(\frac{a^2 G_1}{4b^3} + \frac{(4b^2 - a^2)G_2}{4b^2}\right)(2v_n + v_{n+1} + v_{n-1})$$
$$+ \frac{ah}{2b^2}(\frac{G_1}{b} - G_2)(u_{n-1} - u_{n+1}) - \frac{G_1'}{a}(2v_n - v_{n+2} - v_{n-2}) - M\omega_y^2 v_n \quad (16)$$

### III. DISPERSION RELATION FOR DUST-LATTICE WAVES

In a 2D planar zigzag crystal, longitudinal (for $v_n = 0$) and transverse (for $u_n = 0$) waves can propagate or may cause melting; taking $u_n = u_0 \exp[i(kna/2 - \omega t)]$ and $v_n = 0$ for the longitudinal waves, where the wave number ($k$) obeys $-\pi/2 \leq ka \leq \pi/2$, we obtain the following dispersion relation

$$\omega_L^2 = \omega_x^2 + 4\left(\frac{h^2 G_1}{Mb^3} + \frac{(b^2 - h^2)G_2}{Mb^2}\right)\sin^2(\frac{ka}{4}) + 4\frac{G_2'}{M}\sin^2(\frac{ka}{2}) \quad (17)$$

A similar dispersion relation for transverse waves, can be obtained by assuming $v_n = v_0 \exp[i(kna/2 - \omega t)]$ and $u_n = 0$, namely,

$$\omega_T^2 = \omega_y^2 + \left(\frac{a^2 G_1}{Mb^3} + \frac{(4b^2 - a^2)G_2}{Mb^2}\right)\cos^2(\frac{ka}{4}) + 4\frac{G_1'}{Ma}\sin^2(\frac{ka}{2}) \quad (18)$$

where in equation (17), the term $\left(\frac{h^2 G_1}{Mb^3} + \frac{(b^2 - h^2)G_2}{Mb^2}\right) = G(\kappa)$ is a function of the polynomial coefficients $G_1$ and $G_2$, (i.e., a function of the lattice parameter $\kappa$ or $h$) for a Yukawa system. Fig. (2), shows that for $h_1 = 1/2a$, $G(\kappa)$ is always positive for all values of $\kappa$, however this will not be the case for the different values of "hs", i.e., when $h_2 = \sqrt{3}/2a$, $G(\kappa)$ is negative for $\kappa \leq 1.62$ and positive for $\kappa > 1.62$.

When $h_3 = a$, $G(\kappa)$ is negative for $\kappa \leq 2.44$ and positive for $\kappa > 2.44$. The dependence of $G(\kappa)$ on the value of $h$ can be important and can influence the dispersion relation.
Also choosing different values of "hs" (i.e., $h_1 = 1/2 a$, $h_2 = \sqrt{3}/2 a$, and $h_3 = a$), "βs" (i.e., $\beta_1 = 5.3 \pm .5 \times 10^{-2}$, $\beta_2 = 2.3 \pm 0.3 \times 10^{-2}$ and $\beta_3 = 2.0 \pm .5 \times 10^{-2}$) from Ref. [33] and "κs" helps us study the stability or instability of the structure. For example we can see the effect of $\kappa$ for which $G(\kappa)$ has different signs on the propagations of the waves; such as if for a value of $\kappa$, $\omega^2$ becomes negative (and so $\omega$ becomes complex), then the zigzag configuration becomes unstable, the phase changes, and melting occurs. These results must be in agreement with experimental and simulation results reported previously [21 32-34].

In Fig. (3) we show the longitudinal ($\omega_L$) and the transverse ($\omega_T$) dispersion relations for different values of $\beta$, $h$ and $\kappa$.
We note in Fig. (3) that for large values of $h$ and smaller $\kappa$s (whose $G(\kappa)$ is negative), the longitudinal modes cause the instability of the zigzag structure but when $\kappa$ is larger (for positive $G(\kappa)$), the longitudinal waves can propagate.

Although unlike longitudinal mode, transverse mode behaves differently for any value of $\kappa$ and $h$, hence for this oscillation mode instability is not at all possible and can be observed in Fig. (4).

Now to derive a linear dispersion relation for the coupled DL (dust-lattice) mode, we consider all particle displacements in Eqs. (14) and (16) to be $u_n, v_n \propto \exp[i(kna/2 - \omega t)]$, and obtain

$$\left\{ \omega^2 - \omega_x^2 - 4\left(\frac{h^2 G_1}{Mb^3} + \frac{(b^2 - h^2) G_2}{Mb^2}\right) \sin^2\left(\frac{ka}{4}\right) - 4\frac{G_2'}{M} \sin^2\left(\frac{ka}{2}\right) \right\} \times \left\{ \omega^2 - \omega_y^2 \right.$$
$$\left. - \left(\frac{a^2 G_1}{Mb^3} + \frac{(4b^2 - a^2) G_2}{Mb^2}\right) \cos^2\left(\frac{ka}{4}\right) - 4\frac{G_1'}{Ma} \sin^2\left(\frac{ka}{2}\right) \right\} - \frac{a^2 h^2}{M^2 b^4} \left(\frac{G_1}{b} - G_2\right)^2 \sin^2\left(\frac{ka}{2}\right) = 0 \quad (19)$$

Eq. (19) contains both the longitudinal (x) and transverse vertical (y) DL modes.
Now in order to study the propagation characteristics of the of the dust lattice waves, we carry out numerical analyses of the dispersion relation [given by Eq. (19)]; Figs. (5-7) show the dependence of normalized frequency upon the distance between two chains as a function of normalized wave number both for longitudinal and transverse waves.

In Fig. (5), the distance between two chains is considered to be $h_1$. As mentioned before, when $h$ is small the particle getting closer and a larger confining force must be used to overcome the repulsive electrostatic force. This leads to the result that the acoustic coupled oscillations ($\omega_{ac}$) will not propagate for larger wave numbers. When the distance between two chains increases, the repulsive force in the y direction is very small and the acoustic oscillation mode will be mainly determined by the confinement potential.

Longitudinal and transverse dispersion relations and acoustic and optical coupled branches have been shown for $h_2 = \sqrt{3}/2\, a$ in Figure 6.

As can be noticed in Fig. (6), that the optical coupled branches for small wave numbers and for a small $\kappa$ follow the transverse mode, as the wave number increases they approximate the longitudinal mode. The reverse of this is true for acoustic coupled branches; that is, in the beginning, acoustic coupled branches for small wave numbers behave like longitudinal mode and with the increase of the wave number they approximate the transverse mode. This is not true when $\kappa$ is large. For large values of $\kappa$, the longitudinal mode fits in the acoustic coupled branch and the transverse mode conforms to the optical coupled branch.

The normalized coupling and decoupling branches of modes have been drawn for $h_3 = a$ in Figure 7. In general, the frequencies of the acoustic (optical) branches increase (decrease) with increasing the distance between the two chains.

The behavior of the $\omega_{op}$ frequency mode depends on the distance between the two chains; when it is $h_1$, the $\omega_{op}$ frequency mode is growing and for $h_2$ or $h_3$, it is descending.

## III. CONCLUSIONS

In this work we have presented an analytical model for the planar zigzag crystal of dust particles coupled by the first and second neighbor harmonic forces in 2D.

The amount of constant electric force caused from the electrodes to maintain the zigzag structure so that it could overcome the Yukawa electrostatic force between the particles was also calculated. In doing so, the G (kappa) function was introduced which helped us to determine if the melting occurs or not?

We have discussed the particles having two degrees of freedom, the dispersion relations of longitudinal and transverse dust-lattice modes and the coupling between them due to electrostatic force It was found that not only the decoupling dispersion relations but also the coupling oscillation modes depend on the lattice parameters (like $\kappa$ and $h$). These results are in agreement with experimental and simulations reported previously. For the small $\kappa$, the distance between branches of dispersion relations was obviously large.

Moreover, we have found that the propagation of the $\omega_L$ and $\omega_{ac}$ modes depends on the value of "$h$"; *if it is* smaller than a certain amount, then these modes many not propagate.

This is because when the distance between the two chains is reduced, the repulsive force between them would increase and the frequency of the $\omega_L$ and $\omega_{ac}$ oscillation modes is decreased. Hence, if $h$ is small enough, the longitudinal and acoustic modes will prevail by the restoring force produced by electrodes (i.e., the confinement force), which is why such modes become almost undifferentiated for a small $h$. Furthermore, for small wave numbers and for small $\kappa$ optical coupled branches follow the transverse mode and as the wave number increases they approximate the longitudinal mode. Conversely, for small wave numbers the acoustic coupled branches are similar to the longitudinal mode and with the increase of the wave number they approximate the transverse mode. This does not hold when $\kappa$ is large.

The "analytic relations" presented here, may also be investigated for the nonlinear case in the zigzag structure; however, this is beyond the scope of present work and will be considered in the future.

**Figure Captions:**

Fig. (1): Schematic representation of particles of the zigzag crystal.
Fig. (2): Plot between $G(\kappa)$ and k for three different values of h.
Fig. (3): The normalized solutions of the dispersion relation given by Eq. (17), describing the decoupled modes $\omega_L$ for k = 0, k = 1, k = 2 and k = 3, from left to right, respectively.
Fig. (4): The normalized solutions of the dispersion relation (18), describing the decoupled modes $\omega_T$ for k = 0, 2 and k = 3, from left to right, respectively.
Fig. (5): The normalized acoustic, optical, longitudinal and transverse oscillation modes for k = 0, 2 and k = 3, from left to right, respectively, for $h_1$.
Fig. (6): The normalized coupling and decoupling of modes for k = 0, 2 and k = 3 from left to right, respectively, for $h_2$.
Fig. (7): The normalized coupling between the transverse and longitudinal DL modes and decoupled modes $\omega_L$ and $\omega_T$ are represented for k = 0, 2 and k = 3, from left to right, respectively, for $h_3$.

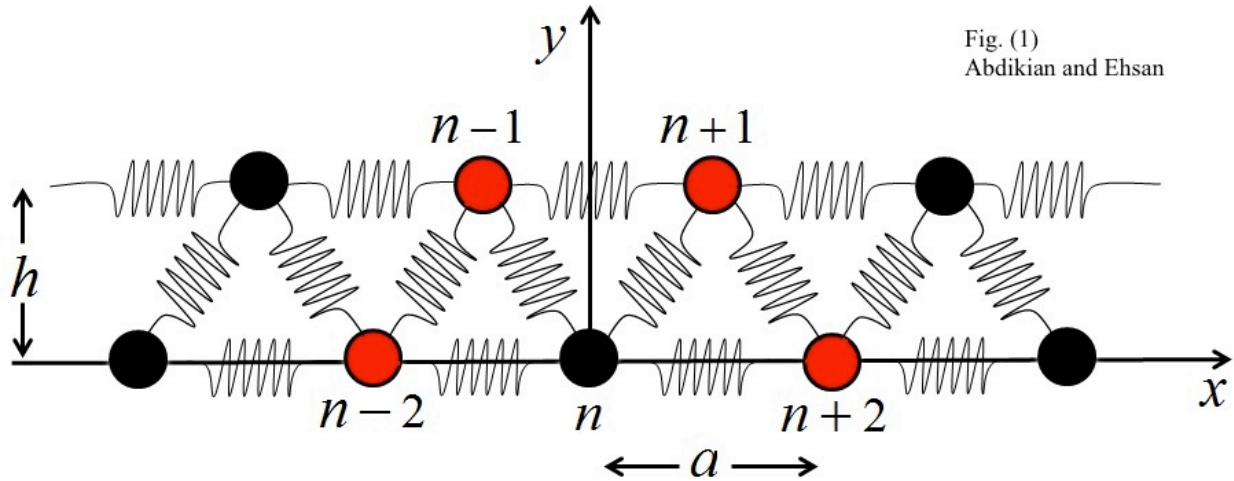

Fig. (1)
Abdikian and Ehsan

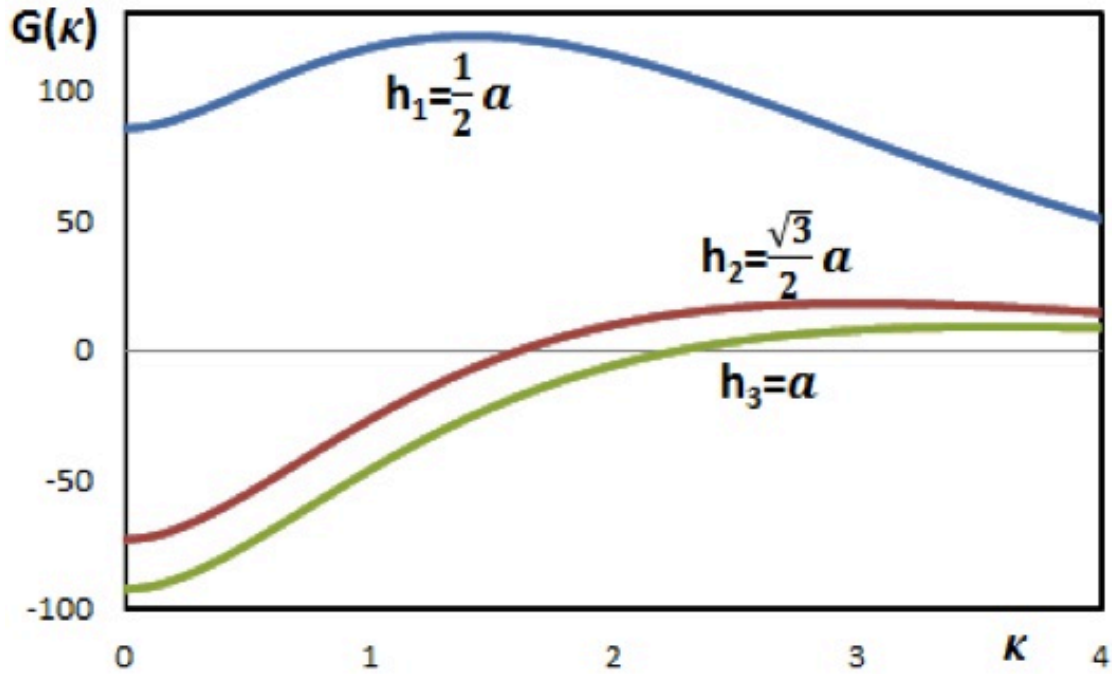

Fig. (2)
Abdikian and Ehsan

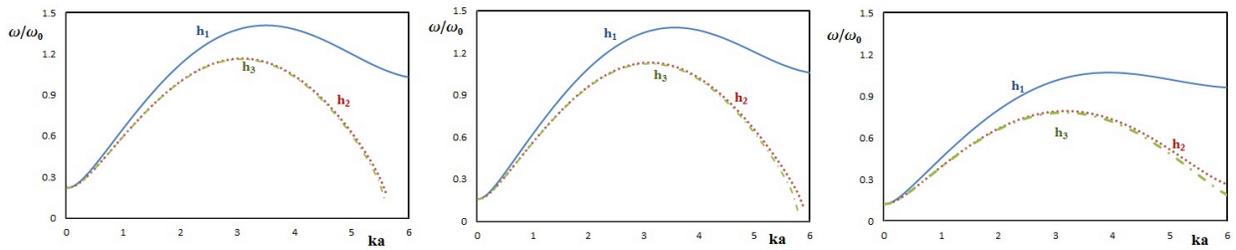

Fig. (3)

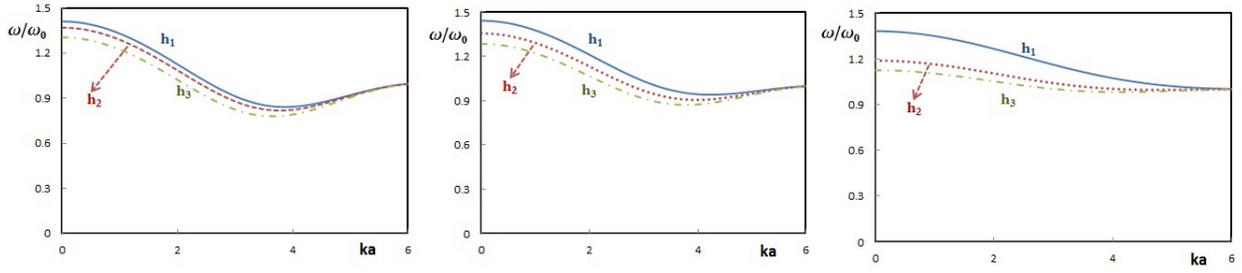
Fig. (4)

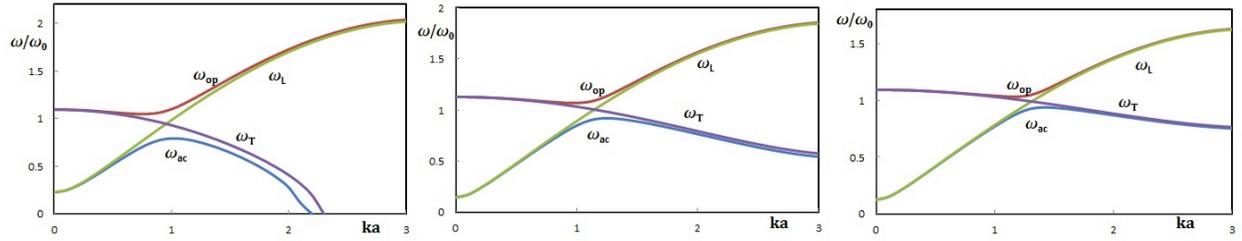
Fig. (5)

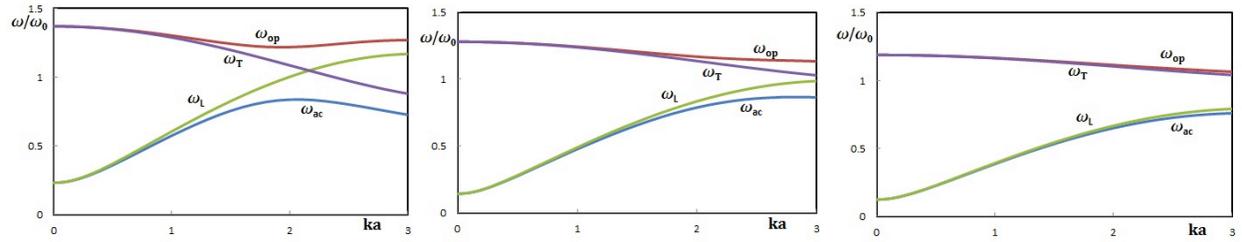
Fig. (6)

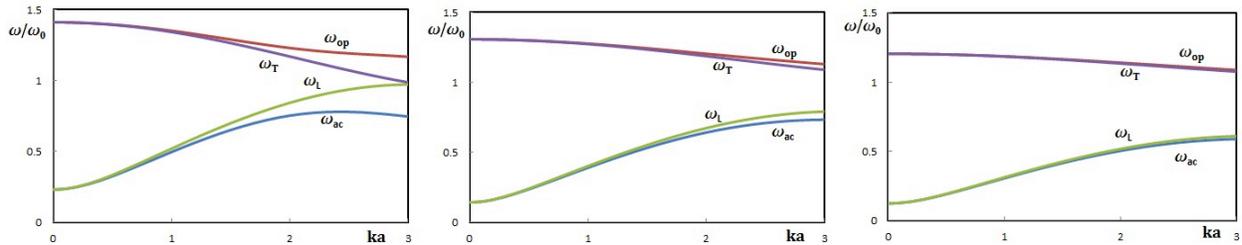
Fig. (7)